\documentstyle[prl,twocolumn,aps]{revtex}
\input{epsf}
\begin{document}
\draft
\twocolumn[\hsize\textwidth\columnwidth\hsize\csname @twocolumnfalse\endcsname
\title{Integrability and Quantum Chaos in Spin Glass Shards}

\author{B.Georgeot and D. L. Shepelyansky$^{(*)}$}

\address {Laboratoire de Physique Quantique, UMR 5626 du CNRS, 
Universit\'e Paul Sabatier, F-31062 Toulouse Cedex 4, France}

\date{July 7, 1998}

\maketitle

\begin{abstract}
We study spin glass clusters (``shards") in a random
transverse magnetic field, and determine the regime where 
quantum chaos and random matrix level statistics emerge
from the integrable limits of weak and strong field.
Relations with quantum phase transition are also discussed.

\end{abstract}
\pacs{PACS numbers: 05.45.+b, 75.10.Jm, 75.10.Nr}
\vskip1pc]

\narrowtext


Quantum manifestations of classical chaos and integrability
have been intensively investigated in the last decade \cite{leshouches}.
It has been realized that the statistical properties of the quantum
energy spectrum are strongly influenced by the underlying classical dynamics.
Indeed,  classical integrability generally implies the absence of
level repulsion and a Poissonian statistics of energy level spacings.
On the contrary,  chaotic dynamics leads to level repulsion with the level
spacing statistics $P(s)$ being the same as in the Random Matrix Theory (RMT),
i.e. the Wigner-Dyson distribution (WD) \cite{bohigas}.   
These two distributions of $P(s)$ also characterize respectively
the localized and
metallic phases in the Anderson model of disordered systems.
At the critical point between these two phases an
intermediate level spacing statistics occurs \cite{shklo}.

While for one-particle systems the statistical properties
of spectra are well understood,  the same problem in many-body systems
has been addressed only recently.   The first results demonstrated 
that many-body integrable systems  are still characterized by Poisson
statistics $P_P(s)$, whereas in absence of integrals of motion
the Wigner-Dyson statistics $P_{WD}(s)$ has been found \cite{bel}.  
More recently, investigations of finite Fermi systems such as the
Ce atom \cite{flam} and the $^{28}$Si nucleus \cite{zel1} put forward
the question of the statistical description and thermalization induced
by interaction.   A quantum chaos criterion for emergence of
RMT statistics and dynamical thermalization
induced by interaction was established in \cite{jac2}. Namely, 
a crossover from
Poisson statistics to the WD distribution takes place when the coupling
matrix elements become comparable to the level spacing between directly 
coupled states.  For two-body interaction, the critical interaction strength
becomes exponentially larger than the multiparticle level spacing.  Indeed
the latter decreases exponentially with the number of particles, while 
the former decreases typically only quadratically.  This result was 
corroborated in \cite{nous},  and should apply 
to various physical systems
such as complex nuclei, atoms, clusters and  quantum dots.

In this Letter we develop and apply these concepts to disordered spin systems.
These systems are of great experimental and theoretical interest \cite{mezard}; in
particular,
quantum spin glasses have recently attracted a great deal of attention and
various approaches have been developed to investigate their properties
\cite{young}.  Such important problems as zero temperature
quantum phase transition, structure of the phase diagram, correlations
and susceptibility properties have been intensively investigated both
analytically and numerically \cite{fisher,huse1,huse2,rieger,grempel,sachdev}.
However, to the best of our knowledge the level spacing statistics has
not been studied in disordered spin systems. Here we present theoretical 
estimates 
for spin glass "shards" which determine
the crossover from integrability to quantum chaos in a way analogous to
the case of finite interacting fermionic systems.  Numerical investigations
of $P(s)$ statistics give a powerful test of this crossover both 
at high temperature $T$ and at $T=0$
where a quantum phase transition is believed to take place.

The spin glass shards we study are described by the Hamiltonian:
\begin{equation}
\label{hamil}
H = \sum_{i<j} J_{ij} \sigma_{i}^x \sigma_{j}^x  + \sum_{i} \Gamma_i
\sigma_{i}^z ,
\end{equation}
where the $\sigma_{i}$ are the Pauli matrices for the spin $i$ and the first
sum runs
over all spin pairs.  The local random magnetic field is represented 
by $\Gamma_i$ uniformly distributed in the interval $[0,\Gamma ]$. 
The exchange interactions $J_{ij}$ are distributed in the same way
in $[-J/\sqrt{n},J/\sqrt{n}]$ where $n$ is the total number of spins.
For $\Gamma=0$ this system is
the classical Sherrington-Kirkpatrick spin glass model \cite{mezard}.
The $\sqrt{n}$ factor in the definition of $J_{ij}$
ensures a proper thermodynamic limit  at $n \rightarrow \infty$.
In the limit $J/\Gamma \rightarrow 0$ one obtains a paramagnetic phase
with all spins in the field direction. The opposite limit ($J/\Gamma
 \rightarrow \infty$) corresponds to a spin glass phase at low temperature
 \cite{mezard}.

Let us first discuss the properties of the model for highly excited states
near the center of the energy band.
At  $J=0$ the system is integrable, since there are $n$ integrals of motion,  
and $P(s)$ should follow the Poisson distribution.  When $J/\Gamma$ increases,
we expect integrability to be destroyed and therefore a crossover 
towards $P_{WD}(s)$ statistics typical of RMT should take place.
At $\Gamma=0$ the model again becomes integrable since there are $n$ operators
($\sigma_i^x$) commuting with the Hamiltonian, hence $P(s)=P_P (s)$.  The
latter may seem surprising, especially in the light of recent discussions
on chaos in classical spin glass \cite{alava}.  However, in spite of a
possible complex thermodynamic behavior (Monte-Carlo dynamics), 
the Hamiltonian dynamics of a classical spin glass is integrable.
As a result, we expect another crossover from a WD statistics back to 
a Poissonian one for large $J/\Gamma$.
This picture is confirmed by the numerical results for $P(s)$ displayed 
in Fig.1. These $P(s)$ distributions were obtained for the states of the same
symmetry class ($S_2$), namely the number of spins up is always an even number 
(the
interaction does not mix states with even and odd number of spins up). We will
call the other symmetry class with an odd number of spins up $S_1$.

\begin{figure}
\epsfxsize=3.4in
\epsfysize=2.6in
\epsffile{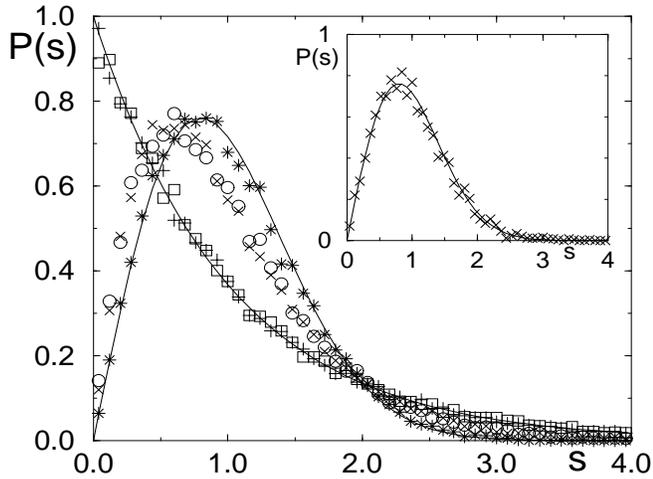}
\vglue 0.2cm
\caption{Crossover from Poisson to WD statistics in the model 
(\ref{hamil}) for the states
in the middle of the energy band ($\pm 12.5 \%$ around the center) for
$n$=12:  $J=0, \eta=0.984$ (+); $J/\Gamma=J_{cp}/\Gamma=0.38, 
\eta=0.3$ ($\times$); 
$J/\Gamma=0.866, \eta=0.027 (*)$; $J/\Gamma=J_{cs}/\Gamma=6.15, \eta=0.3$ (o);
 $\Gamma=0, \eta=0.99$ ($\Box$). Full curves
 show the Poisson and WD distributions.  Total statistics ($NS$) is more than 
 $3 \times 10^4$; $s$ is in units of mean level spacing.
 Insert shows $P(s)$ for the first excitation from the ground state in the 
 chaotic regime for $n=15$, $J/\Gamma=0.465$, $\eta_0=0.018$ ($NS=3000$) 
($\times$); 
 the full line shows $P_{WD}(s)$. } 
\label{fig1}
\end{figure}

To analyze the evolution of the $P(s)$ distributions with respect to the 
coupling $J$, it is convenient to use the parameter 
$\eta=\int_0^{s_0}
(P(s)-P_{WD}(s)) ds / \int_0^{s_0} (P_{P}(s)-P_{WD}(s)) ds$,
where  $s_0=0.4729...$ is the intersection point of $P_P(s)$ and $P_{WD}(s)$
\cite{jac2}.  In this way $\eta=1$
corresponds to the Poissonian case, and $\eta$=0 to the WD distribution.
The data of Fig.1 show that the distributions with the same value of $\eta$ 
are very close, even if $J/\Gamma$ varies more than ten times.
It is convenient to determine a critical coupling strength $J_c$ at which
the crossover from $P_P(s)$ to $P_{WD}(s)$ takes place by the condition
$\eta(J_c/\Gamma)=0.3$.  
The variation of $\eta$ with respect to $J$ is shown in Fig.2 for different
$n$.  The global behavior of $\eta$ is in agreement with the
above picture of transition between integrability and quantum chaos.  Indeed,
when $J/\Gamma$ increases, $\eta$ drops to zero and then starts to grow
again back to one when the spin glass term in (\ref{hamil}) dominates.
The fact that a WD statistics sets in is not trivial.  Indeed, in dimension
$d=1$ (the model (\ref{hamil}) with nearest-neighbor coupling only
and $J_{i i+1}$ drawn randomly from [$0,J$], which was studied in
\cite{fisher,young}) the statistics remains Poissonian ($\eta=1$)
 at arbitrary $J/\Gamma$, as is shown in Fig.2.  The critical
 point $J/\Gamma=1$ \cite{fisher} does not manifest itself. 
The physical reason for this
behavior is based on the fact that this $d=1$ model can be mapped into
a model of non-interacting fermions (see e. g. \cite{fisher}). Therefore
the total energy is the sum of one-particle energies that generically leads
to $P_P(s)$.  As a result RMT can never be applied to this model
and dynamical (interaction induced) thermalization never takes place.
Thermalization can only appear through a coupling to a thermal bath that
was implicitly used in \cite{yosa,sachdev}.
On the contrary in the system (\ref{hamil}) there are two transitions from
integrability to chaos which determine two critical couplings $J_{cp}$, from
the paramagnetic side, and $J_{cs}$ from the spin glass side 
($\eta(J_{cs})=\eta(J_{cp})=0.3; J_{cs}> J_{cp}$). 
\vskip -0.5cm
\begin{figure}
\epsfxsize=3.4in
\epsfysize=2.6in
\epsffile{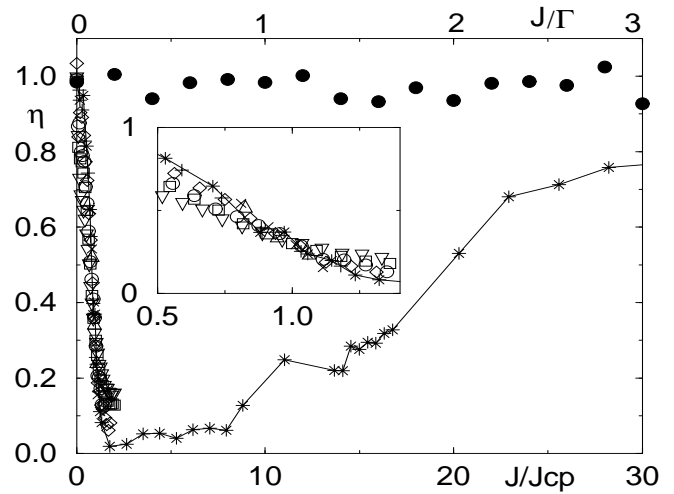}
\vglue 0.2cm
\caption{Dependence of $\eta$ on the rescaled coupling strength $J/J_{cp}$
for the states in the middle of the energy band  for
$n=7 (\bigtriangledown),8 (\Box),9 $(o)$,10 (\diamond),11 (+),12 (*),13 (\times),
14  (\bigtriangleup)$ and symmetry $S_2$; $NS$ varies between $12500$ and $160000$.
 The full line  for $n=12$ 
shows the global behavior.  Data for the $d=1$ model (see text)
 are given for $n=12$ (full circles) as a function of $J/\Gamma$ (upper scale);
$NS=30000$.
The insert magnifies the region near $J/J_{cp}=1$.} 
\label{fig2}
\end{figure}

In analogy with finite interacting fermionic systems \cite{jac2} we expect
that quantum chaos sets in when the coupling strength $U$
becomes comparable
to the spacing between directly coupled states $\Delta_c$ ($U\sim \Delta_c$).
For small $J$
in the middle of the spectrum one has $U\sim J/\sqrt{n}$ and 
$\Delta_c \sim 16\Gamma/ n^2 $ since each state is coupled to $ n(n-1)/2$
states in an energy band of $8\Gamma$.  This gives the quantum chaos border
from the paramagnetic side:
\begin{equation}
\label{jcp}
J_{cp} \approx C_p \Gamma /n^{3/2},
\end{equation}
where $C_p$ is some numerical constant. We note that $J_{cp}$ is exponentially
larger than the multiparticle spacing $\Delta_n \sim 2 n \Gamma /2^n$.
For large $J$ the transitions between unperturbed states at $J \gg \Gamma$
are determined by the $\Gamma$ term in (\ref{hamil}), and have typical
value $\Gamma$.  The number of such transitions, in a typical energy 
interval $J$, is $n$, hence $\Delta_c \sim J/n$.  This gives the quantum chaos
border from the spin glass side:
\begin{equation}
\label{jcs}
J_{cs} \approx C_s \Gamma n,
\end{equation}
where again $C_s$ is some constant.

These theoretical predictions are confirmed by the numerical results
presented in Figs.2-3 which give $C_p\approx 16$ and $C_s \approx 0.5$.  
We attribute the deviation from (\ref{jcp}) in Fig.3 seen for small $n$   
to the fact that for these values $J_{cp}$ is
rather large; this can slightly 
modify the unperturbed spectrum and the estimates for $\Delta_c$.  Also
for small $n$ the proximity of the second transition at $J_{cs}$ can  
affect the actual value of $J_{cp}$.  The data
shown in the insert of Fig.2 indicate that $\eta$ depends only on the
ratio $J/J_{cp}$; the global scaling behavior is less visible
than in the fermionic model \cite{jac2}, apparently due to the reasons above.

\begin{figure}
\epsfxsize=3.4in
\epsfysize=2.6in
\epsffile{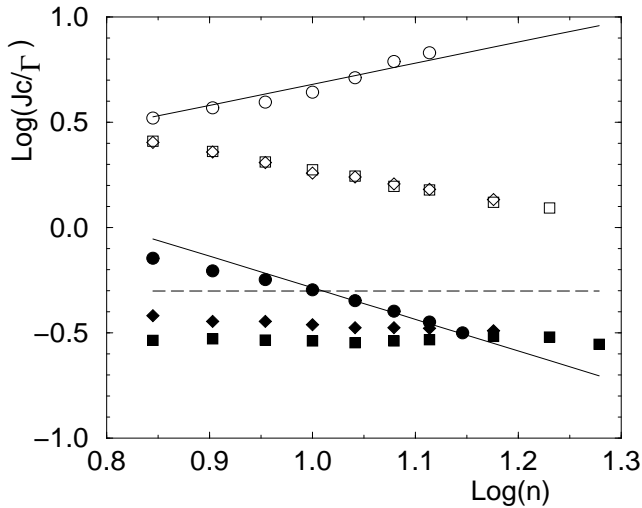}
\vglue 0.2cm
\caption{Critical coupling strength $J_c$ as a function of $n$:
$J_{cp}$ at band center for symmetry $S_2$ (full circles), and near the
ground state (full squares: $S_1$; full diamonds: $S_2$ ); $J_{cs}$ at 
band center for $S_2$ (o), and near 
the ground state ($\Box$: $S_1$; $\diamond$: $S_2$).  
Full lines show the theory (Eqs. (\ref{jcp}),
(\ref{jcs})) for the band center; the dashed line indicates the quantum
critical point at $T=0$, $J/\Gamma=0.5$.  Logarithms are decimal.} 
\label{fig3}
\end{figure}

The analysis above at the band center corresponds to a very high energy or
temperature.  Therefore the properties of (\ref{hamil}) near the ground state
energy
$E_g$ should be studied separately.  To do that we investigated $P(s)$ for the
first excitations from $E_g$ within the same symmetry class.
For the class $S_1$, the distribution $P(s)$ was computed 
for the first four level
spacings starting from $E_g$ by averaging over different disorder realizations. 
The first excitation
and its $\eta$ value ($\eta_0$) was analyzed separately while the other
three were used to generate one cumulative $P(s)$ with an $\eta$ noted 
by $\eta_1$ \cite{sym}.  There are several physical reasons for this separation:
in the presence of a gap $\eta_0$ is related to the gap fluctuations while
$\eta_1$ characterizes the quasi-particle excitations; in a 
metallic quantum dot
with non-interacting electrons $\eta_0$ reflects the chaotic character
of the dot giving  $\eta_0 =0$ while the spectrum of higher excitations 
becomes close to $P_P(s)$ with $\eta_1 \approx 1$ due to
the absence of interaction. 

\begin{figure}
\epsfxsize=3.4in
\epsfysize=2.6in
\epsffile{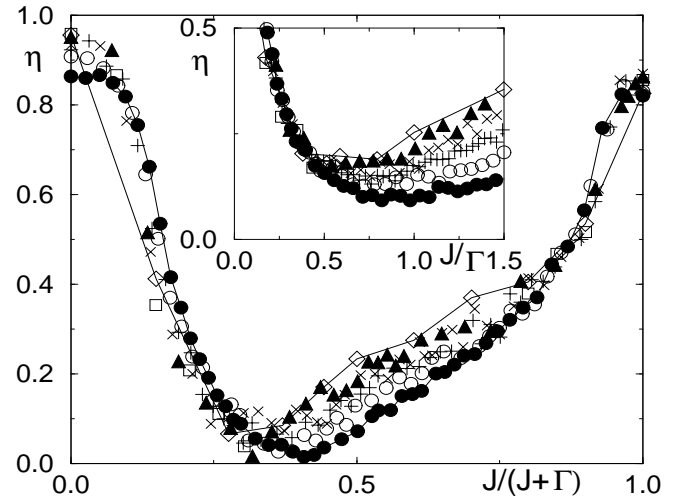}
\vglue 0.2cm
\caption{Dependence of $\eta_0$ on $\tilde{J}=J/(J+\Gamma)$:
$n=7$ (full circles), $n=9 $(o)$, n=11 (+), n=13 (\times), 
n=15$ (full triangles), $n=17 (\diamond), n=19 (\Box)$;
 $2000 \leq NS \leq 30000$. Full curves connect data
for $n=7,17$.
Insert shows $\eta_1$ in the region near $J/\Gamma=J_q/\Gamma\approx 0.5$
 in more detail; $6000 \leq NS \leq
90000$.  Data are for $S_1$ symmetry.} 
\label{fig4}
\end{figure}

The global behavior of $\eta_0$ and $\eta_1$ as a function of
$\tilde{J}=J/(J+\Gamma)$ is shown in Fig.4.  This parametrization naturally
represents the two integrable limits $J=0$ and $\Gamma=0$ in a symmetric way.
In both extreme cases ($\tilde{J}=0;1$),
$\eta_0$ and $\eta_1$ approach the Poissonian value
$1$.  In between there is a pronounced minimum with $\eta_0 \approx 0$ and
$\eta_1 \approx 0.1$.  When the number of spins in the shard
increases,  the values of $\eta$ to the right from the minimum go up
markedly, whereas the minimal $\eta$ value itself changes only weakly.
  These $\eta$ data suggest
the existence of a crossing point around $J =J_q \approx 0.5 \Gamma$.
  For $J< J_q$ the difference
between curves for different $n$ is small, whereas
it is much bigger for $J>J_q$.
   The data in Fig.3 show the variation of $J_{cp}, J_{cs}$
(for which $\eta_1=0.3$) with $n$; the behavior of $J_{cs}$
is consistent with $J_{cs} \rightarrow J_q$ for $n \rightarrow \infty$
 for both symmetry classes.  
Also according to these data 
$J_{cp} \propto n^{-\alpha}$ with $0 \leq \alpha \leq 0.2$.   This value
of $\alpha$ is smaller than the one expected from the quantum chaos 
criterion, used to derive (\ref{jcp}).  Indeed, for small $J$ near the ground
state $\Delta_c \sim \Gamma/n$ whereas the coupling is still $J/\sqrt{n}$, and
therefore one expects $\alpha=0.5$.  For large $n$ this would imply that 
$\eta$ drops quickly to zero with $J$ 
and then increases again after the crossing
point $J_q$. In such a scenario, in the thermodynamic limit 
($n \rightarrow \infty$) we expect $J_{cp} \rightarrow 0$ and
$J_{cs} \rightarrow J_q$, so that
$\eta$ changes from zero ($J < J_q$) to one ($J>J_q$) with some intermediate
statistics  at the
critical point $J_q$.  The determination of the $\eta$ value
at that point for $n \rightarrow \infty$ requires larger system sizes.  
We expect that this critical point corresponds
to the quantum phase transition between
paramagnetic and spin glass phases  discussed in \cite{young,huse1,grempel} 
for a case when all $\Gamma_i = \Gamma$.  The above scenario is similar to
the situation for the Anderson transition in $d=3$ \cite{shklo}: 
as for the Anderson insulator,  the lack of space ergodicity in 
the spin glass phase implies $\eta=1$.
However, in the paramagnetic phase the situation
may be more complicated.  Indeed, the criterion $U\sim \Delta_c$ which
gives $\alpha=0.5$ assumes that the couplings remain small and do not
strongly modify the unperturbed $\Delta_c$.  But near $E_g$ the typical
mean field acting on a spin is of the order $J$ and for 
$J \approx \Gamma/\sqrt{n}$ is much larger than the
local fields $\Gamma_i \sim \Gamma/n$.  The strong mean field near $E_g$
can change the effective $\Delta_c$.  Such
a fact was seen for the Ce atom \cite{flam}.  This may lead to another
scenario in which in the thermodynamic limit 
$J_{cp}$ tends to some non-zero value smaller than $J_q$.
Such a behavior would be in agreement with the small variation 
of $J_{cp}$  with $n$ in Fig.3.
\vskip -0.5cm
\begin{figure}
\epsfxsize=3.4in
\epsfysize=2.6in
\epsffile{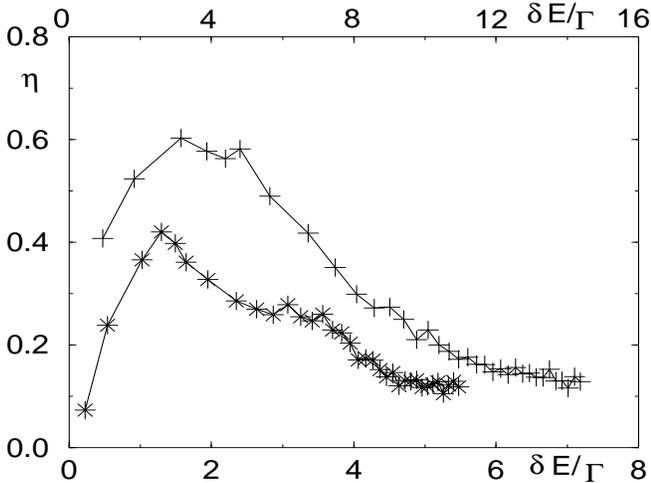}
\vglue 0.2cm
\caption{Dependence of $\eta$ on $\delta E /\Gamma$ for $n=12$
($1000$ realizations of disorder):
$J/\Gamma= 0.52 (*) $(lower scale), $J/\Gamma=3.46 (+)$ (upper scale).
Data are for $S_1$ symmetry.} 
\label{fig5}
\end{figure}

The discussions above dealt with the two extreme cases of energy values.
The variation of $\eta$ between these two limits is shown in Fig.5 near
the critical point and in the
spin glass phase as a function of the
excitation energy $\delta E=E-E_g$.  
Both cases show an unusual behavior when $\eta$ initially grows with energy
and starts to decrease only later.  This tendency is more pronounced
near the critical point.
A possible reason for this behavior is that the relative influence of
the mean field becomes weaker as $\delta E / \Gamma$ increases.  It is also
possible that the situation is similar to the case of a chaotic quantum dot with 
noninteracting electrons discussed above: the ground state is chaotic
but the interaction between the first quasi-particle excitations is weak
and so initially $\eta$ grows with $\delta E$.  Larger system sizes 
are required to clarify this issue.

In conclusion, our study shows an unusual and interesting
interplay between integrability, quantum chaos and phase 
transitions in disordered spin systems.

We thank F. Mila and E. S{\o}rensen for stimulating discussions.


\vskip -0.5cm
\begin{thebibliography}{99}
\bibitem[*]{byline1} Also Budker Institute of Nuclear Physics,
630090 Novosibirsk, Russia
\bibitem{leshouches} {\it Les Houches Lecture Series} {\bf 52},
        Eds. M.-J. Giannoni, A.Voros, and J. Zinn-Justin (North-Holland,
        Amsterdam, 1991).
\bibitem{bohigas} O. Bohigas, M.-J. Giannoni, and
        C.Schmit, Phys. Rev. Lett. {\bf 52}, 1 (1984); 
        O.Bohigas in \cite{leshouches}.
\bibitem{shklo} B. I. Shklovskii, B. Shapiro, B. R. Sears, P. Lambrianides and
H. B. Shore, Phys. Rev. B {\bf 47}, 11487 (1993); D.Braun, G.Montambaux and
        M.Pascaud, cond-mat/9801099.
\bibitem{bel} G. Montambaux, D. Poilblanc, J. Bellissard and C. Sire, Phys.
Rev. Lett. {\bf 70}, 497 (1993); D. Poilblanc, T. Ziman, J. Bellissard, F.
Mila and G. Montambaux, Europhys. Lett. {\bf 22}, 537 (1993).
\bibitem{flam} V. V. Flambaum, A. A. Gribakina, G. F. Gribakin and M. G.
Kozlov, Phys. Rev. A {\bf 50}, 267 (1994);
V. V. Flambaum, A. A. Gribakina, G. F. Gribakin and I. V.
Ponomarev, Phys. Rev. E., {\bf 57} 4933 (1998).
\bibitem{zel1} V. Zelevinsky, B. A. Brown, N. Frazier and M. Horoi, Phys. Rep.
{\bf 276}, 85 (1996).
\bibitem{jac2} P.~Jacquod and D.~L.~Shepelyansky, Phys. Rev. Lett. {\bf 79},
                 1837 (1997).
\bibitem{nous} B. Georgeot and D.~L.~Shepelyansky, Phys. Rev. Lett. {\bf 79},
4365 (1997); D.Weinmann, J.-L.Pichard and Y.Imry,  J. Phys. I France
{\bf 7}, 1559 (1997); A. D. Mirlin and Y. V. Fyodorov, 
Phys. Rev. B {\bf 56}, 13393 (1997); R. Berkovits and Y. Avishai, 
Phys. Rev. Lett. {\bf 80}, 568 (1998).
\bibitem{mezard} K.Binder and A.P.Young, Rev. Mod. Phys.
{\bf 58}, 801 (1986); M.M\'ezard, G.Parisi and M.A.Virasoro,
{\it Spin Glass Theory and Beyond}, World Scientific, Singapore (1987).
\bibitem{young} {\it Spin Glasses and Random Fields}, Ed. A.P.Young,
World Scientific, Singapore (1997).
\bibitem{fisher} D. S. Fisher, Phys. Rev. Lett. {\bf 69}, 534 (1992);
Phys. Rev. B {\bf 51} 6411 (1995).
\bibitem{huse1} J. Miller and D. A. Huse, Phys. Rev. Lett. {\bf 70}, 3147
(1993).
\bibitem{huse2} M. Guo, R. N. Bhatt and D. A. Huse, Phys. Rev. Lett. {\bf 72},
4137 (1994); Phys. Rev. B {\bf 54} 3336 (1996).
\bibitem{rieger} H. Rieger and A. P. Young, Phys. Rev. Lett. {\bf 72}, 4141
(1994); Phys. Rev. B {\bf 54} 3328 (1996). 
\bibitem{grempel} D. R. Grempel and M. J. Rozenberg, Phys. Rev. Lett. 
{\bf 80}, 389 (1998); cond-mat/9802106.
\bibitem{sachdev} S. Sachdev, cond-mat/9705266.
\bibitem{alava} M. Alava and H. Rieger, cond-mat/9804136.
\bibitem{yosa} S.~Sachdev and A.~P.~Young, Phys. Rev. Lett. {\bf 78}, 2220
(1997).
\bibitem{sym} In the $S_2$ symmetry class, $\eta_0 \approx 0.4$ at $J=0$;
this complicates the analysis  of  $ \eta_0$, but the results 
for $\eta_1$ in Fig.3 are similar to the ones of $S_1$.
\end{thebibliography}
\end{document}